\documentclass[amssymb,preprint, onecolumn,floats,amsmath,showpacs,superscriptaddress]{revtex4}

\usepackage{bm}
\usepackage{graphicx}
\usepackage{amssymb}
\usepackage{amsfonts}
\usepackage{amsmath}
\usepackage{epstopdf}

\def \FUW{Institute of Experimental Physics, Faculty of Physics, University of Warsaw, ul. Pasteura 5, 02-093 Warsaw, Poland}
\def \IT{National Institute of Telecommunication, ul. Szachowa 1, PL-04-894 Warsaw, Poland}

\begin{document}

\title{Angle Dependence of Photonic Enhancement of Magneto-Optical Kerr Effect in DMS Layers}

\author{M. Koba} \affiliation{\FUW, \IT}
\author{J.~Suf\mbox{}fczy\'{n}ski} \affiliation{\FUW}
\email{Jan.Suffczynski@fuw.edu.pl}


\begin{abstract}
We  investigate theoretically an angle dependence of enhancement of polar magneto-optical Kerr effect (MOKE) obtained thanks to a deposition of a paramagnetic Diluted Magnetic Semiconductor (DMS) layer on one-dimensional photonic crystal layer. Our transfer matrix method based calculations conducted for TE and TM polarizations of the incident light predict up to an order of magnitude stronger MOKE for a (Ga,Fe)N DMS layer when implementing the proposed design. The maximum enhancement for TE and TM polarization occurs for the light incidence at the normal and at the Brewster angle, respectively. This indicates a possibility of tuning of the MOKE enhancement by adjustment of the polarization and the incidence angle of the light.
\end{abstract}

\maketitle

\section{Introduction}
The magneto-optical Kerr effect (MOKE)\cite{Kerr}, manifested by a rotation of a linear polarization plane of the light reflected from the magnetized material's surface, provides an excellent tool for probing of metals\cite{Erskine:PRL1973} or Dilute Magnetic Semiconductors (DMS)\cite{Furdyna-JAP-1988, Cibert:book08} magnetization. The DMSs have been intensively studied recently\cite{Dietl-Science-2000, Gamelin:PRL2005, Ando-science-2006, Kuroda:NM2007, Dietl-NM-2010} due to a wide range of valuable functionalities resulting from a combination of properties inherent for a semiconductor and for a magnetic material. In particular, weak losses typical for a semiconductor host and a strong magnetic circular dichroism (MCD) specific to ferromagnetic metals is profitable for implementation of the DMS in optical isolators or spatial light modulators in advanced spintronic or photonic applications. Alike the MCD, the MOKE results generally from a difference in the absorption of the left- and the right-circularly polarized light in the magnetized material. In the case of the DMS, the MOKE in the fundamental gap spectral region originates from a splitting of interband transitions in magnetic field induced by spin-orbit and $s,p$-$d$ exchange interactions between band carriers and localized spins of the magnetic dopant ions.

Practical applications in magneto-optical devices necessitate strong magneto-optical effects from possibly thin DMS films. This could be, in principle, achieved through raising of a concentration of a magnetic dopant, \textit{x} \cite{Furdyna-JAP-1988, Gaj:1978, Gaj:1979}. However, optical performance of the DMSs typically drops down with the increasing \textit{x} \cite{Furdyna-JAP-1988, Pacuski:PRB2011}. As recently shown, magneto-optical response of DMS can be efficiently enhanced while keeping \textit{x} at a reasonably low level by boosting the light-matter interaction exploiting photonic \cite{Gourdon-SSC-2002, Sadowski:PRB1987, Shimizu:APL2001, Inui:JMMM2011} or plasmonic \cite{Ganshina:JPCM2010} effects. In particular, a significant enhancement of a Faraday rotation or MCD related to (Ga,Mn)As layers has been obtained thanks to embedding a DMS layer in a resonant microcavity formed by two Distributed Bragg Reflectors (DBR) \cite{Shimizu:APL2001, Inui:JMMM2011}.

Here, we investigate theoretically angle dependencies in a photonic enhancement of the MOKE in a paramagnetic (Ga,Fe)N. The enhancement is obtained thanks to a deposition of (Ga,Fe)N layer on the top of a single, (Al,Ga)N based, DBR layer. Through a systematic study we determine first a design of the structure assuring the highest degree of the MOKE enhancement in the fundamental gap spectral region. We find a strong impact of the Fabry-Perot light interferences on the MOKE magnitude, in contrary to the case of optically thick magnetized metal layers, where interference effects are precluded by a small light penetration depth. Next, we analyze a dependence of the MOKE in the DMS and of its enhancement on the light incidence angle. The angle dependencies of the MOKE has been studied so far only for magnetized metallic films \cite{You:APL1996, Grychtol:PRB2010}. We show that with the proposed design, the MOKE enhancement for both, transverse electric (TE) and transverse magnetic (TM), polarizations remains insensitive to the angle of the light incidence in a wide angle range. However, the maximum of MOKE enhancement for TM polarization is obtained for the Brewster angle, similarly as in the case of magnetized metal layers \cite{You:APL1996, Grychtol:PRB2010}.

The possibility of enhancement of the MOKE in (Ge,Fe)N is especially appealing in a view of wide ranging applications foresee~\cite{Dietl-Science-2000, Dietl-NM-2010} for a whole class of wide-band gap DMSs~\cite{Gamelin:PRL2005, Pacuski:book10, Pacuski:PRB2011, Suffczynski:PRB2011}. The photonic approach to the MOKE enhancement presented here can be, however, easily applied to any, not only the wide band gap, DMS.

\section{Structures}
\label{sec:structure}
\begin{figure}
\includegraphics[width=0.8\columnwidth]{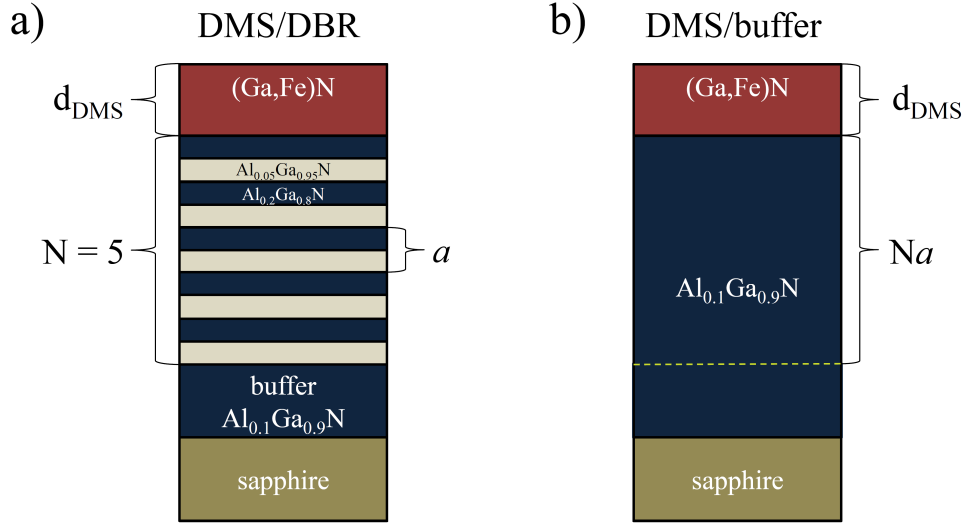}
\caption{
(Colour on-line) a) DMS/DBR structure: DMS layer (paramagnetic Ga$_{0.998}$Fe$_{0.002}$N) of thickness d$_{\rm{DMS}}$ deposited on Al$_{0.05}$Ga$_{0.95}$N/Al$_{0.2}$Ga$_{0.8}$N DBR mirror (N = 5 periods, each period of a thickness $a$), grown on a Al$_{0.1}$Ga$_{0.9}$N buffer. b) A reference, DMS/buffer, structure: the same DMS layer deposited on a Al$_{0.1}$Ga$_{0.9}$N buffer of a thickness corresponding to the Bragg mirror thickness in the DMS/DBR structure.
\label{fig:structure_schemes}}
\end{figure}

We consider a magnetooptical response of two structures: one involving a DMS layer deposited on a DBR layer (referred to as 'DMS/DBR'), and the other one ('DMS/buffer') serving as a reference, involving the DMS layer deposited on a buffer layer (see fig. \ref{fig:structure_schemes}). The DMS/DBR structure consists of the paramagnetic (Ga,Fe)N layer ($x_{Fe}$=0.2\%) of thickness d$_{\rm{DMS}}$, the Bragg mirror constituted by $N = 5$ periods of alternating Al$_{0.05}$Ga$_{0.95}$N/Al$_{0.2}$Ga$_{0.8}$N layers, and Al$_{0.1}$Ga$_{0.9}$N buffer (thickness $d_{{buf\!fer}} = 200$~nm) deposited on a sapphire substrate. The assumed relatively small number of periods and a low difference of Al content in the layers within the DBR mirror is enough for a significant MOKE enhancement in DMS~\cite{jemwa:mkjs}. At the same time, such design is advantageous from the point of view of an epitaxial growth of a good quality (with a low dislocations density) structures~\cite{Shenk-APL-2002}.
In the reference DMS/buffer structure, the DBR layer is replaced by the Al$_{0.1}$Ga$_{0.9}$N buffer layer of a thickness equal to the overall thickness of the DBR layer, as shown in fig.~\ref{fig:structure_schemes}b).

\section {Model}
\label{sec:model}
In this work, we consider the case of polar MOKE, where the magnetization vector is perpendicular to the sample reflection surface and parallel to a plane of the light incidence. The angle between the linear polarization direction of the incident light and the major axis of the elliptical polarization of the reflected light defines the Kerr rotation angle $\Theta_K$, given by (e.g., Ref. \cite{Testelin}):
\begin{equation}
\Theta_K=\frac{1}{2}\arg\frac{r^-}{r^+},
\label{eq:Theta_K}
\end{equation}
where ${r^-}$ and ${r^+}$ represent complex,  wavelength dependent amplitudes of reflection coefficients for ${\sigma^-}$ and ${\sigma^+}$ circular polarizations, respectively.
\begin{figure}
\includegraphics[width=0.8\columnwidth]{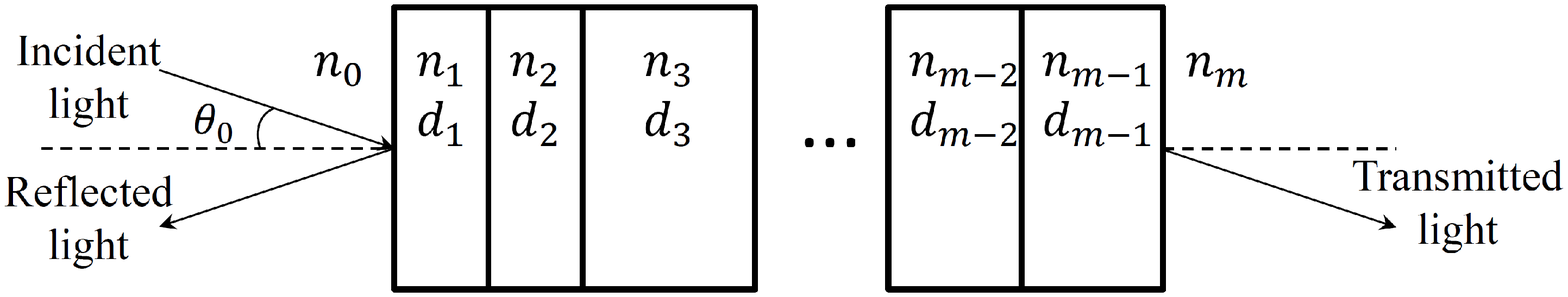}
\caption{Multilayered structure considered in the calculations of reflection coefficient $r$. The $\theta_0$ denotes the angle of the light incidence.
\label{fig:tmm_schem}}
\end{figure}
We start with a determination of a reflection coefficient $r$ in a general case of an arbitrary stack of dielectric layers (see fig.~\ref{fig:tmm_schem}). We assume that all layers constituting the structure shown in fig.~\ref{fig:tmm_schem} are uniform, isotropic, and infinite in the plane parallel to the sample surface. The $i$th layer of thickness $d_i$ is described by a wavelength dependent, complex refractive index $n_i (\lambda)$. The $n_i (\lambda)$ is calculated as a square root of a dielectric function $\varepsilon_i (\lambda)$ for $i$th layer taking into account contributions from interband optical transitions to the absorption. Namely, transitions to discrete ground and excited excitonic states with a Lorentzian line shape, as well as to a continuum of unbound states~\cite{Tanguy:PRL1995, Pacuski:PRL2008} are taken into account. Sice we deal with a wurtzite type semiconductor, where the valence band at k = 0 is split into three subbands, the $A$, $B$, and $C$ excitonic states are included to the model~\cite{Dingle:PRB1971, GilBriot:PRB1995, Stepniewski:PRB1997}. The same value of a background dielectric constant $\varepsilon^{\ast}_0$ = 5.2 is assumed for all GaN-based layers~\cite{Stepniewski:PRB1997}. Polaritonic effects~\cite{Pacuski:PRL2008} are neglected~\cite{JGR:PRB2013}. Parameters of excitonic transitions (i.e., energy positions, linewidths and oscillator strengths) in the DMS layer as a function of magnetic field for ${\sigma^-}$ and ${\sigma^+}$ circular polarizations of the light are taken from reflectivity experiment performed at T = 2 K on Ga$_{1-x}$Fe$_x$N ($x_{Fe} = 0.2$~\%) layer grown on a GaN buffer (fig. 4 in Ref.~\onlinecite{JGR:PRB2013}). With a direct link to the experiment, our simulations are expected to describe properly a real structure performance. In particular, a contribution to MOKE originating from variation of not only energy, but also a shape (i.e., oscillator strength and linewidth) of the excitonic transitions in magnetic field is taken into account. Variation of all three parameters describing the excitonic transition has been shown recently to contribute with a comparable weight to a Faraday rotation~\cite{Maslana:2001} and to the MCD~\cite{JGR:PRB2013} effects.

We consider TE and TM polarizations of the light incident on the interface between the first (0) and the second (1) layer at an arbitrary angle $\theta_0$ (see fig.~\ref{fig:tmm_schem}). Within the transfer matrix method~\cite{Macleod} formalism the tangential electric (\textit{E}) and magnetic (\textit{H}) fields at the first ($0/1$) and the last ($(m-1)/m$) interface are related by:

\begin{eqnarray}
& \left[ \begin{array}{c} E_{0/1} \\ H_{0/1} \end{array} \right] = M \left[ \begin{array}{c} E_{(m-1)/m} \\ H_{(m-1)/m} \end{array} \right] &,
\end{eqnarray}

The characteristic matrix of the structure $M$ is defined as $M =\left[ \small\begin{array}{cc} M_{11} & M_{12} \\ M_{21} & M_{22}\end{array}\right] = \prod_{i=1}^{m-1}M_i$, where matrix M$_i$ is defined for the $i$th layer as:
\begin{equation}
M_i=\left[ \begin{array}{cc} \cos{\beta_i} & (-i \sin{\beta_i})/q_i \\ -iq_i \sin{\beta_i} & \cos{\beta_i}\end{array} \right],
\end{equation}
where
\begin{equation}
\beta_i=\frac{2\pi d_i}{\lambda} {\left(n_i^2-n_0^2\sin^2{\theta_0}\right)^{1/2}},
\end{equation}
and
\begin{equation}
q_i=\frac{\left(n_i^2-n_0^2\sin^2{\theta_0}\right)^{1/2}}{n_i^2}
\label{eq:tmm_q_tm}
\end{equation}
for the TM polarization, while
\begin{equation}
q_i={\left(n_i^2-n_0^2\sin^2{\theta_0}\right)^{1/2}}
\label{eq:tmm_q_te}
\end{equation}
for the TE polarization.
The reflection coefficient $r$ is finally obtained as:
\begin{equation}
r=\frac{(M_{11}+M_{12}q_m)q_0-(M_{21}+M_{22}q_m)} {(M_{11}+M_{12}q_m)q_0+(M_{21}+M_{22}q_m)}.
\label{eq:tmm_r}
\end{equation}

The $r$ for TM or TE polarizations is obtained by an appropriate substitution of equation \ref{eq:tmm_q_tm} or \ref{eq:tmm_q_te} into equation \ref{eq:tmm_r}, respectively. Reflectivity coefficients ${r^-}$ and ${r^+}$ for the particular case of our structures are obtained from the eq.~\ref{eq:tmm_r} solved with excitonic parameters respective for a given magnetic field and circular polarization of the light.

In the initial step of our investigation of the angle dependence of photonic MOKE enhancement we consider the normal incidence case and determine a design of the structure providing the strongest MOKE related to the DMS layer. We calculate the wavelength dependent Kerr rotation angle $\Theta_K (\lambda)$ in (Ga,Fe)N fundamental band gap spectral region for the DMS/DBR and DMS/buffer structures described in previous Section. We vary the DMS layer thicknesses $d_{\rm{DMS}}$ between 50 nm and 350 nm. The central wavelength of the mirror $\lambda_{\rm{DBR}}$ is tuned between 345 nm and 365 nm by tuning thicknesses $d_i$ of the layers constituting the DBR so that the condition $d_i = \lambda_{\rm{DBR}}/(4n_i)$ is fulfilled. Once the $\Theta_K (\lambda)$ is determined, the MOKE magnitude is calculated as an integral of the absolute value of the $\Theta_K (\lambda)$ over the wavelengths within the region of (Ga,Fe)N excitonic transitions. The integral describes properly the MCD~\cite{JGR:PRB2013} or the MOKE magnitude in the case when neighboring excitons of finite linewidths contribute to the dielectric function. In addition, a difference between extrema of the MOKE curve in the excitonic region (\textit{peak to peak} amplitude) is also determined from $\Theta_K (\lambda)$ curve.


\section{Numerical results}\label{sec:numerical_results}
\begin{figure}
\includegraphics[width=0.5\linewidth]{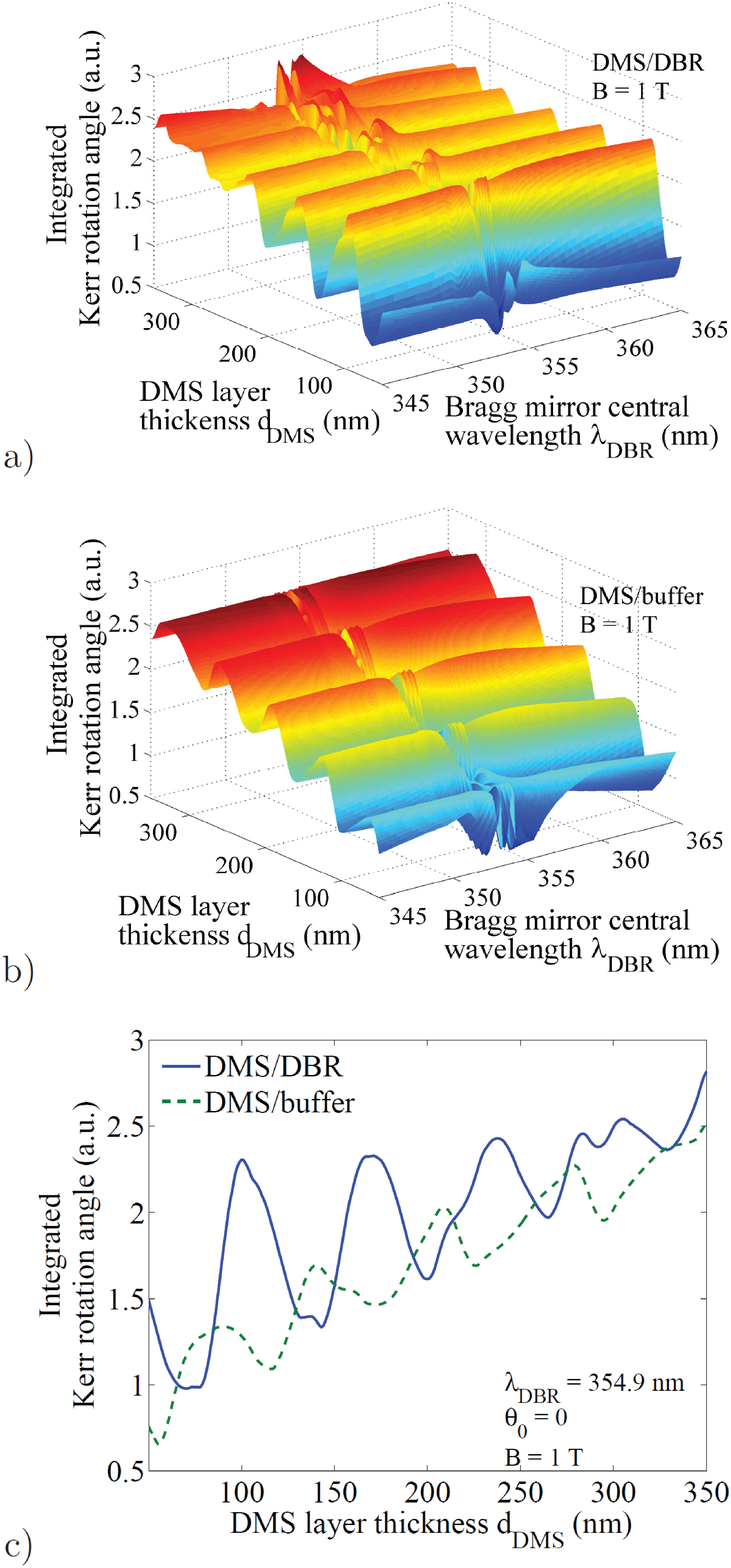}
\caption{(Colour on-line) Calculated magnitude of the MOKE versus the DBR mirror central wavelength $\lambda_{\rm{DBR}}$ and the DMS layer thickness $d_{\rm{DMS}}$ at B = 1 T for (a) the designed DMS/DBR structure and (b) the reference DMS/buffer structure in the normal light incidence case ($\theta_0$=0). The variation of the $\lambda_{\rm{DBR}}$ in b) refers to the variation of the Al$_{0.1}$Ga$_{0.9}$N buffer thickness so that it is equal to a total thickness of the DBR it replaces. (c) A cross-section of the plots shown in a) and b) at $\lambda_{\rm{DBR}} = 354.9$.
\label{fig:integral_dbr_bulk}}
\end{figure}

The calculated magnitude of the MOKE for the normal incidence case ($\theta_0$=0) versus the Bragg mirror central wavelength $\lambda_{\rm{DBR}}$ and the DMS thickness $d_{\rm{DMS}}$ is depicted in fig.~\ref{fig:integral_dbr_bulk}a) and fig.~\ref{fig:integral_dbr_bulk}b) for DMS/DBR and DMS/buffer structures, respectively. Magnetic field is fixed at B = 1 T, corresponding to the magnetization of the sample close to its saturation value~\cite{Pacuski:PRL2008, JGR:PRB2013}. A comparison of figs. \ref{fig:integral_dbr_bulk}a) and \ref{fig:integral_dbr_bulk}b) allows for a discussion of the impact of the sample design on its magnetooptical response.
First, fig.~\ref{fig:integral_dbr_bulk} shows that the MOKE magnitude generally increases with the increasing DMS layer thickness. The apparent oscillatory character of the dependence results from the light interference effects. In the case of the DMS/DBR structure (fig.~\ref{fig:integral_dbr_bulk} a)), boundaries accounting for the interferences are constituted by the DBR mirror and a DMS/air interface. For the d$_{\rm{DMS}}$ corresponding approximately to a multiple of $\lambda_{\rm{DBR}}/(2n_{\rm{DMS}})$, the partial reflections sum up in phase and the light interferes constructively, accounting for a multiple passage of the light through the DMS layer. This enhances the absorption and the resulting MOKE magnitude. The oscillations fade with the increasing $d_{\rm{DMS}}$ due to a saturation of the light absorption within the DMS layer.
The interference effects are less pronounced in the case of the DMS/buffer structure, as it is well seen in in fig.~\ref{fig:integral_dbr_bulk} c) showing a cross-section of the plots figs.~\ref{fig:integral_dbr_bulk} a) and \ref{fig:integral_dbr_bulk} b) at $\lambda_{\rm{DBR}} = 354.9$. Indeed, due to a relatively small difference of refractive indices, the reflectivity coefficient of (Ga,Fe)N/(Al,Ga)N interface is an order of magnitude smaller than the one of the DBR mirror in the DMS/DBR structure.

The second, fig.~\ref{fig:integral_dbr_bulk} b) indicates that the MOKE integral is particularly sensitive to the $\lambda_{\rm{DBR}}$ when DBR stop band overlaps spectrally with the excitonic transitions. The highest MOKE magnitude is obtained at around $\lambda_{\rm{DBR}}$ = 355 nm, where both, real and imaginary, parts of the DMS refractive index exhibit sharp extrema due to a contribution to the dielectric function coming from excitonic absorption peaks~\cite{Tanguy:PRL1995, Stepniewski:PRB1997, JGR:PRB2013}. As seen in fig. \ref{fig:integral_dbr_bulk} c), thanks to the implementation of the resonant DBR, a substantial (around two fold) enhancement of MOKE integral yet for thin ($\sim$ 100 nm) semimagnetic layers is obtained.

Impact of the sample design is typically not taken into account when discussing the MOKE or other magneto-optical phenomena in observed in experiments on DMS layers. The above discussion shows that MOKE magnitude depends not only on the properties of the DMS layer itself, but also on the design of the whole structure involving the DMS layer. Our calculations indicate thus a possibility of interference enhancement of the MOKE magnitude even in the structures with no photonic crystal. Moreover, a difference between the MOKE in DMSs (where the light penetrates the structure) and the MOKE in optically thick metallic layers (where the light probes mainly the surface) is clearly highlighted here.

\begin{figure}
\includegraphics[width=0.5\columnwidth]{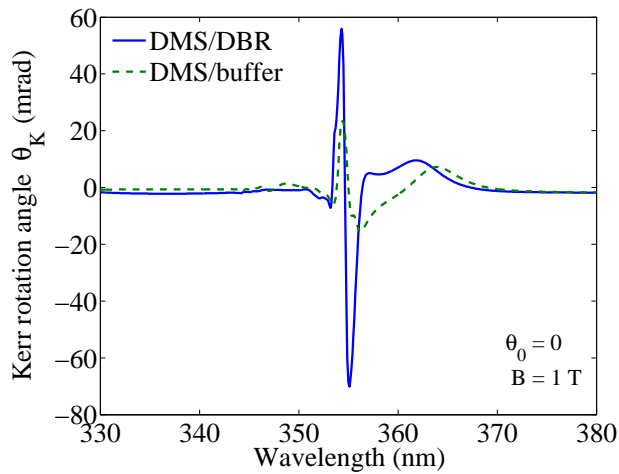}
\caption{
(Colour on-line) Kerr rotation angle $\Theta_K (\lambda)$ at B = 1 T versus the light wavelength, for the DMS/DBR (blue solid line) and DMS/buffer (green dashed line) structures in the case of the normal light incidence ($\theta_0=0$).
\label{fig:Kerr_plot_0}}
\end{figure}

After a demonstration of the photonic enhancement of the MOKE in the normal light incidence case, we now pass to an analysis of the oblique incidence angles case. We set the (Ga,Fe)N layer thickness to $d_{\rm{DMS}} = 100$ nm. The thicknesses of the layers within the DBR are set equal to, respectively, 27.6 nm and 33.1 nm for Al$_{0.05}$Ga$_{0.95}$N and Al$_{0.2}$Ga$_{0.8}$N layers, yielding the $\lambda_{\rm{DBR}} = 354.9$ nm. This assures that the DBR mirror stopband overlaps spectrally with the $A$ and $B$ excitons. To analyze an evolution of the Kerr rotation angle $\Theta_K (\lambda)$ curve with the angle of incidence $\theta_0$ we first present in fig.~\ref{fig:Kerr_plot_0} the $\Theta_K (\lambda)$ for $\theta_0=0$ at B = 1 T for both, DMS/DBR and DMS/buffer, structures. The MOKE amplitude much larger in the case of the DMS/DBR than the DMS/buffer structure and a peak-to-peak enhancement reaching a factor of $\sim$ 3 is evidenced.

We note that despite that two spectrally close (Ga,Fe)N excitons, $A$ and $B$, contribute to the MOKE, the  $\Theta_K (\lambda)$ curves exhibit a Lorentzian-like shape, similar as it would be in the case of a single excitonic transition~\cite{Testelin, Gaj:1978}. (The exciton $C$ of a small oscillator strength~\cite{JGR:PRB2013} is neglected in the discussion.) It can be understood taking into account a non-zero energy difference between $A$ and $B$ excitons at B = 0 T and their opposite splitting in magnetic field. In such a case their antisymmetrical, Lorentzian type contributions~\cite{Testelin, Maslana:2001} to the MOKE do not cancel out, but rather sum up, similarly as in the case of the excitonic MCD in wurtzite structure (Ga,Fe)N~\cite{JGR:PRB2013}.
\begin{figure}
\includegraphics[width=0.5\columnwidth]{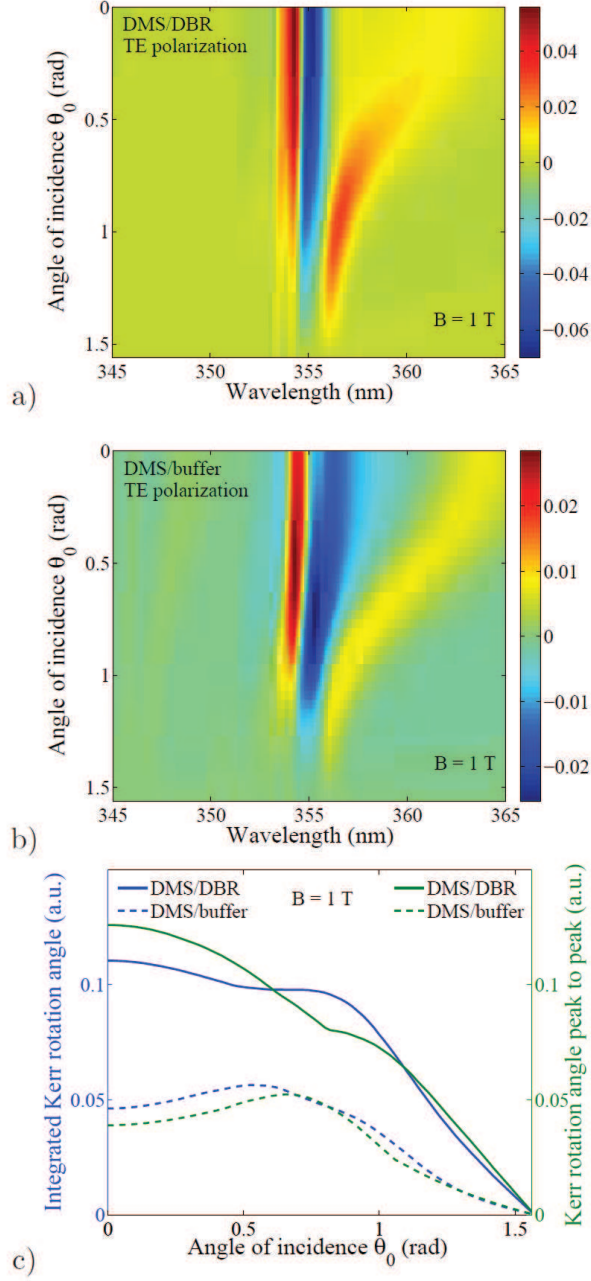}
\caption{
(Colour on-line) Calculated Kerr rotation angle $\Theta_K (\lambda)$ versus $\theta_0$ of the TE polarized light at B = 1 T, for (a) DMS/DBR and (b) DMS/buffer structure. (c) Left axis: integral of $\Theta_K (\lambda)$ over the wavelengths in the excitonic region as a function of $\theta_0$ for DMS/DBR (thick blue solid line) and DMS/buffer (thick blue dashed line) structures. Right axis: peak-to-peak amplitude of $\Theta_K (\lambda)$ in the excitonic region as a function of $\theta_0$ for DMS/DBR (thin green solid line) and DMS/buffer (thin green dashed line).
\label{fig:Kerr_mesh_TE}}
\end{figure}
The figs.~\ref{fig:Kerr_mesh_TE}(a) and \ref{fig:Kerr_mesh_TE}(b) show the $\Theta_K (\lambda)$ curves as a function of the wavelength and the angle of incidence $\theta_0$ of TE polarized light for the DMS/DBR and DMS/buffer structures, respectively. It is seen in figs.~\ref{fig:Kerr_mesh_TE}(a) and (b) that the spectral shape of the MOKE curve for both structures practically does not change with the $\theta_0$ within the full range of the incidence angles. Resulting integrals of the absolute value of MOKE, as well as a peak to peak amplitude of $\Theta_K (\lambda)$ in the excitonic spectral region are plotted versus $\theta_0$ in fig.~\ref{fig:Kerr_mesh_TE}(c). The fig.~\ref{fig:Kerr_mesh_TE}(c) indicates that for both structures the MOKE magnitude (independently how parameterized) keeps approximately a constant value for the $\theta_0$ from 0 to $\sim$0.8 rad ($\sim$46 deg), it drops considerably for larger angles and, as expected, reaches 0 for $\theta_0 \rightarrow \frac{\pi}{2}$ rad (90 deg). However, in the whole range of the incidence angles the magnitude of the MOKE determined for the DMS/BBR structure remains much largee as compared to the case of the DMS/buffer structure. This confirms the advantage of the implementation of 1-D photonic crystal for the enhancement of the MOKE related to (Ga,Fe)N.
\begin{figure}
\includegraphics[width=0.5\columnwidth]{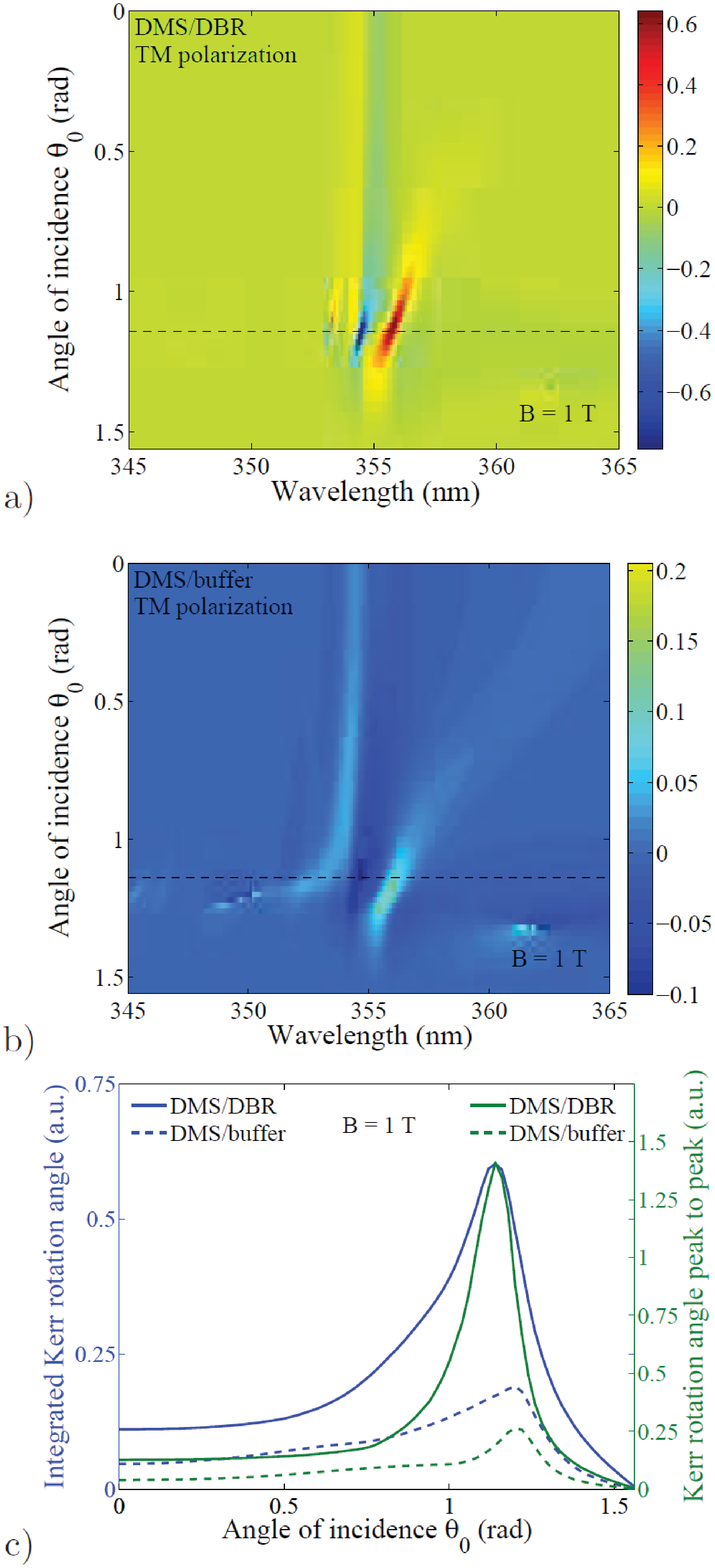}
\caption{
(Colour on-line) The same is in fig. \ref{fig:Kerr_mesh_TE}, but for TM polarized light. The dashed line in a) and b) indicates the Brewster angle at $\theta_0 \sim$1.14 rad ($\sim$65.3 deg).
\label{fig:Kerr_mesh_TM}}
\end{figure}
Fig.~\ref{fig:Kerr_mesh_TM} shows an analogous set of plots such as in fig.~\ref{fig:Kerr_mesh_TE}, but for the TM polarization of the incident light. A comparison of either integrated MOKE curves (blue lines in fig.~\ref{fig:Kerr_mesh_TM}(c)) or peak-to-peak amplitudes (green lines in fig.~\ref{fig:Kerr_mesh_TM}(c)), indicates that also in the case of the TM polarization around two fold MOKE enhancement is obtained for angles below $\theta_0 \sim$0.8 rad ($\sim$46 deg) thanks to the implementation of the photonic structure. However, a strong maximum, not present in the case of the TE polarized light is observed at around $\theta_0 \sim$1.14 rad ($\sim$65.3 deg). This angle corresponds to the Brewster (polarization) angle for the TM polarized light incident on air/(Ga,Fe)N layer interface. A lossless light transmission through the air/dielectric interface, conjuncted with a high Bragg mirror reflectivity in the DMS/DBR structure, leads to a strong phase dependence of the light on the (Ga,Fe)N refractive index. As a result, an order of magnitude enhanced MOKE is obtained. The enhanced MOKE at the Brewster angle incidence has been so far observed only for magnetized metal films~\cite{You:APL1996, Grychtol:PRB2010}. Above result implies that through adjustment of the light polarization and the incidence angle in the vicinity of the Brewster angle one can effectively control the degree of the MOKE enhancement in the DMS.
\section{Summary}
\label{sec:summary}
The obtained results show that magnitude of the MOKE in DMS is not defined solely by the concentration of a magnetic dopant in the DMS layer, as typically assumed, but it depends in a substantial degree also on the actual sample design. It is established that a proper interpretation of the absolute MOKE magnitude in DMS layers requires taking into account effects of the light interference in a semiconductor structure. The implementation of the DBR layer leads to a significant enhancement of the MOKE in the DMS, which is maintained for a wide range of the light incidence angles. The maximum (an order of magnitude) enhancement of the MOKE is obtained for the TM polarized light incident at Brewster angle. This indicates a possibility of effective tuning of the degree of the MOKE enhancement through an adjustment of the light incidence angle and polarization.
\acknowledgments
We thank Tomasz Dietl for valuable discussions. The work was supported by the ERC through the FunDMS Advanced Grant (\#227690) within the ''Ideas'' 7th Framework Programme of the EC, by Polish NCN projects: DEC-2011/01/B/ST3/02406, DEC-2011/02/A/ST3/00131, UMO-2012/05/B/ST7/02155 and by Polish NCBiR project LIDER.


\begin{thebibliography}{99}
\bibitem{Kerr} J.~Kerr, ``On Rotation of the Plane of the Polarization by Reflection from the Pole of a Magnet," {\it Philosophical Magazine Series 5\/}, Vol.~3, No.~19, 321--343, 1877.

\bibitem{Erskine:PRL1973} Erskine, J. and Stern, E., "Magneto-optic Kerr Effect in Ni, Co, and Fe," {\it Phys. Rev. Lett.\/}, Vol. 30, 1329--1332, 1973.

\bibitem{Furdyna-JAP-1988} Furdyna,~J.~K., ``Diluted magnetic semiconductors," {\it J. Appl. Phys.\/}, Vol.~64, No.~4, R29--R64, 1998.

\bibitem{Cibert:book08} Cibert J. \and Scalbert D., "Spin Physics in Semiconductors," {\it Springer Series in Solid-State Sciences, Springer, Heidelberg, ed. Dyakonov M. I.\/}, 389, 2008.

\bibitem{Dietl-Science-2000} Dietl,~T., Ohno,~H., Matsukura,~F., Cibert,~J., and Ferrand,~D., ``Zener {Model} {Description} of {Ferromagnetism} in {Zinc-Blende} {Magnetic} {Semiconductors}," {\it Science\/}, Vol.~287, No.~5455, 1019--1022, 2000.

\bibitem{Dietl-NM-2010} Dietl,~T., ``A ten-year perspective on dilute magnetic semiconductors and oxides," {\it Nature Materials\/}, Vol.~9, No.~12,
    965--974, 2010.

\bibitem{Gamelin:PRL2005} Kittilstved K. R.,  Norberg N. S., and  Gamelin D. R., " "Chemical Manipulation of High-TC Ferromagnetism in ZnO Diluted Magnetic Semiconductors," {\it Phys. Rev. Lett.\/}, Vol.~94, 147209, 2005.

\bibitem{Ando-science-2006} Ando,~K., ``Seeking Room-Temperature Ferromagnetic Semiconductors," {\it Science\/}, Vol.~312, No.~5782, 1883--1885, 2006.

\bibitem{Kuroda:NM2007} Kuroda S., Nishizawa N., Takita K., Mitome M., Bando Y., Osuch K. and Dietl T., "Origin and control of high-temperature ferromagnetism in semiconductors," {\it Nat. Mater. \/}, Vol.~6, 440, 2007.

\bibitem{Gaj:1978} Gaj,~J.~A., Ga{\l}\c{a}zka,~R.~R., and Nawrocki,~M., ``Giant exciton Faraday rotation in {Cd$_{1-x}$Mn$_x$Te} mixed crystals," {\it Solid State Commun.\/}, Vol.~25, No.~3, 193--195, 1978.

\bibitem{Gaj:1979} Gaj,~J.~A., Planel,~R., and Fishman,~G., ``Relation of magneto-optical properties of free excitons to spin alignment of {Mn$^{2+}$} ions in {Cd$_{1−x}$Mn$_x$Te}," {\it Solid State Commun.\/}, Vol.~29, No.~5, 435--438, 1979.

\bibitem{Pacuski:PRB2011} Pacuski,~W., Suf\mbox{}fczy\'nski,~J., Osewski,~P., Kossacki,~P., Golnik,~A., Gaj,~J.~A., Deparis,~C., Morhain,~C., Chikoidze,~E., Dumont,~Y., Ferrand,~D., Cibert,~J., Dietl,~T., "Influence of s,p-d and s-p exchange couplings on exciton splitting in Zn$_{1-x}$Mn$_{x}$O," {\it Phys. Rev. B\/}, Vol.~84, No.~3, 035214 1--7, 2011.

\bibitem{Gourdon-SSC-2002} Gourdon,~C., Lazard,~G., Jeudy,~V., Testelin,~C., Ivchenko,~E.~L., and Karczewski,~G., ``Enhanced {Faraday} rotation in {CdMnTe} quantum wells embedded in an optical cavity," {\it Solid State Commun.\/}, Vol.~123, No.~6--7, 299--304, 2002.

\bibitem{Sadowski:PRB1987} Sadowski,~J., Mariette,~H., Wasiela,~A., Andre,~R., Merle d'Aubigne,~Y., and Dietl,~T., ``Magnetic tuning in excitonic {Bragg} structures of {(Cd,Mn)Te/(Cd,Zn,Mg)Te}," {\it Phys. Rev. B\/}, Vol.~56, No.~4, R1664--R1667, 1997.

\bibitem{Shimizu:APL2001} Shimizu, H., Miyamura, M. and Tanaka, M., ``Magneto-optical properties of a GaAs:MnAs hybrid structure sandwiched by GaAs/AlAs distributed Bragg reflectors: Enhanced magneto-optical effect and theoretical analysis," {\it Appl. Phys. Lett.\/}, Vol.~78, No.~11, 1523-1525, 2001.

\bibitem{Inui:JMMM2011} Inui,~C., Ozaki,~S., Kura ~H. and Sato, T., "Enhancement of Faraday effect in one-dimensional magneto-optical photonic crystal including a magnetic layer with wavelength dependent off-diagonal elements of dielectric constant tensor," {\it J. Magn. Magn. Mater.\/}, Vol.~323, No.~18�19, 2348 - 2354, 2011.

\bibitem{Ganshina:JPCM2010} Ganshina E. A., Golik L. L., Kovalev V. I., Kunkova Z. E., Temiryazev A. G., Danilov Yu. A., Vikhrova O. V., Zvonkov B. N., Rubacheva A. D., Tcherbak P. N., Vinogradov A. N. \and Zhigalina O. M., "Resonant enhancement of the transversal Kerr effect in the InMnAs layers," {\it J. Phys.: Condens. Matter\/}, Vol.~22, 396002, 2010.

\bibitem{You:APL1996} You, Ch-Y and Shin, S-Ch., "Derivation of simplified analytic formulae for magneto-optical Kerr effects," {\it Appl. Phys. Lett.\/}, Vol.~69, 1315, 1996.

\bibitem{Grychtol:PRB2010} Grychtol, P. and Adam, R. and Valencia, S. and Cramm, S. and B\"urgler, D. E. and Schneider, C. M., "Resonant magnetic reflectivity in the extreme ultraviolet spectral range: Interlayer-coupled Co/Si/Ni/Fe multilayer system," {\it Phys. Rev. B\/}, Vol.~82, 054433, 2010.

\bibitem{Pacuski:book10} Pacuski,~W., "Optical spectroscopy of wide-gap diluted magnetic semiconductors," {\it Introduction to the Physics of Diluted Magnetic Semiconductors, Springer Series in Materials Science\/}, Springer, Heidelberg, 2010.

\bibitem{Suffczynski:PRB2011} Suf\mbox{}fczy\'nski,~J., Grois,~A., Pacuski,~W., Golnik,~A., Gaj,~J.~A., Navarro-Quezada,~A., Faina,~B., Devillers,~T. and Bonanni,~A., "Effects of s,p-d and s-p exchange interactions probed by exciton magnetospectroscopy in {(Ga,Mn)N}," {\it Phys. Rev. B\/}, Vol.~83, No.~9, 094421 1--8, 2011.

\bibitem{jemwa:mkjs} Koba,~M., and Suf\mbox{}fczy\'nski,~J.,``Magneto-optical effects enhancement in DMS layers utilizing {1-D} photonic crystal," {\it J. Electromagnet. Wave\/}, Vol.~27, No.~6, 700--706, 2013.

\bibitem{Shenk-APL-2002} Schenk H. P. D., de Mierry P., Vennegues P., Tottereau O., Laugt M., Vaille M., Feltin E., Beaumont B., Gibart P., Fernandez S., and Calle F., "In situ growth monitoring of distributed GaN-AlGaN Bragg reflectors by metalorganic vapor phase epitaxy," {\it Appl. Phys. Lett\/}, Vol.~80, No.~2, 174--176, 2002.

\bibitem{Testelin} Testelin,~C., and Rigaux,~C., ``Resonant magneto-optic {Kerr} effect in {CdTe/Cd$_{1-x}$Mn$_x$Te} quantum-well structures," {\it Phys. Rev. B\/}, Vol.~55, No.~4, 2360--2367, 1997.

\bibitem{Pacuski:PRL2008} Pacuski,~W., Kossacki,~P., Ferrand,~D., Golnik,~A., Cibert,~A., Wegscheider,~M., Navarro-Quezada,~A., Bonanni,~A., Kiecana,~M., Sawicki,~M., and Dietl,~T., ``Observation of Strong-Coupling Effects in a Diluted Magnetic Semiconductor {Ga$_{1-x}$Fe$_x$N}," {\it Phys. Rev. Lett.\/}, Vol.~100, No.~3, 037204 1--4, 2008.

\bibitem{Tanguy:PRL1995} Tanguy, C., "Optical Dispersion by Wannier Excitons," {\it Phys. Rev. Lett.\/}, Vol. 75, No. 22, 4090 1--3, 1995.

\bibitem{Dingle:PRB1971} Dingle, R., Sell, D. D., Stokowski, S. E., Dean, P. J., and Zetterstrom, R. B., "Absorption, Reflectance, and Luminescence of GaN Single Crystals," {\it Phys. Rev. B\/}, Vol.~3, No.~2, 497--500, 1971.

\bibitem{GilBriot:PRB1995} Gil,~B., Briot,~O., and Aulombard,~R.-L., ``Valence-band physics and the optical properties of {GaN} epilayers grown onto sapphire with wurtzite symmetry," {\it Phys. Rev. B\/}, Vol.~52, No.~24, R17028--R17031, 1995.

\bibitem{Stepniewski:PRB1997} St\c{e}pniewski,~R., Korona,~K.~P., Wysmo\l{}ek,~A., Baranowski,~J.~M., Paku\l{}a,~K., Potemski,~M., Martinez,~G., Grzegory,~I., and Porowski,~S., ``Polariton effects in reflectance and emission spectra of homoepitaxial {GaN}," {\it Phys. Rev. B\/}, Vol.~56, No.~23, 15151--15156, 1997.

\bibitem{JGR:PRB2013} Rousset,~J.-G., Papierska,~J., Pacuski,~W., Golnik,~A., Nawrocki,~M., Stefanowicz,~W., Stefanowicz,~S., Sawicki,~M., Jakie\l{}a,~R., Dietl,~T., Navarro-Quezada,~A., Faina,~B., Li,~T., Bonanni,~A., and Suf\mbox{}fczy\'nski,~J., ``Relation between exciton splittings, magnetic circular dichroism, and magnetization in wurtzite (Ga,Fe)N", {\it Phys. Rev. B\/}, Vol.~88, No.~11, 115208--115215, 2013.

\bibitem{Maslana:2001} Ma\ifmmode \acute{s}\else \'{s}\fi{}lana,~W., Mac,~W., Gaj,~J.~A., Kossacki,~P., Golnik,~A., Cibert,~J., Tatarenko,~S., Wojtowicz,~T., Karczewski,~G., and Kossut,~J., ``Faraday rotation in a study of charged excitons in Cd$_{1-x}$Mn$_{x}$Te," {\it Phys. Rev. B \/}, Vol.~63, No.~16, 165318 1--7, 2001.

\bibitem{Macleod} Macleod,~H.~A., {\it Thin-Film Optical Filters \/}, Institute of Physics Publishing, Bristol and Philadelphia, 2001.

\end{thebibliography}
\end{document}